\documentclass[aps,prd,superscriptaddress,showpacs,tighten,nofootinbib]{revtex4}
\usepackage{epsfig}
\usepackage{amssymb}
\usepackage{rotate}
\usepackage{float}
\usepackage{graphicx}
\usepackage{slashed}
\usepackage{amsmath}
\usepackage{setspace}
\usepackage{slashed}
\usepackage{color}
\usepackage{comment}
\usepackage{dcolumn}   
\usepackage{bm}        
\usepackage{multirow}
\newcommand{\ba}{\begin{array}}
\newcommand{\ea}{\end{array}}
\newcommand{\bd}{\begin{displaymath}}
\newcommand{\ed}{\end{displaymath}}
\newcommand{\be}{\begin{equation}}
\newcommand{\ee}{\end{equation}}
\newcommand{\bea}{\begin{eqnarray}}
\newcommand{\eea}{\end{eqnarray}}

\def\th13 {\theta_{13}}

\usepackage{graphicx}

\newcommand{\mathsym}[1]{{}}
\newcommand{\unicode}[1]{{}}

\catcode`\@=11
\def\lsim{\mathrel{\mathpalette\@versim<}}
\def\gsim{\mathrel{\mathpalette\@versim>}}
\def\@versim#1#2{\vcenter{\offinterlineskip
\ialign{$\m@th#1\hfil##\hfil$\crcr#2\crcr\sim\crcr } }}
\catcode`\@=12

\catcode`@=12
\begin{document}

\title{The cosmological constant from the zero point energy of compact dimensions }

\author{Vikram Soni}
\email{vsoni.physics@gmail.com}
\affiliation{Centre for Theoretical Physics, Jamia Millia Islamia, \\ Jamia Nagar, New Delhi-110025, India}

\author{Naresh Dadhich}
\email{nkd@iucaa.ernet.in}
\affiliation{Centre for Theoretical Physics, Jamia Millia Islamia, \\ Jamia Nagar, New Delhi-110025, India}
\affiliation{IUCAA, Pune- 411007, India}

\author{Rathin Adhikari}
\email{rathin@ctp-jamia.res.in}
\affiliation{Centre for Theoretical Physics, Jamia Millia Islamia, \\ Jamia Nagar, New Delhi-110025, India}

\begin{abstract}

We consider extra compact dimensions as the origin of a cosmological universal energy density in the regular dimensions, with  only graviton  fields propagating in
 the compact space dimensions. The quantum zero point energy originating from the finite size boundary condition in the compact dimensions can produce a constant 
energy density in regular $3$ space which is homogeneous and isotropic. It then makes a natural identification with the cosmological constant in conformity with the Einstein equation. It turns out that for the emergent energy density to agree with the observed value of the cosmological constant, the size/radius of compact dimension is to be of order of $10^{-2}$ cm.

\end{abstract}
\pacs{ 98.80.Es,98.65.Dx,98.62.Sb}
\maketitle
\newpage

\section{Introduction}
 
The cosmological constant, $\Lambda$, has strong observational support from the supernova observations \cite{super}. 
The stress tensor of  quantum field theoretic vacuum energy  turns out to be, like the cosmological constant, proportional to metric. 
This is how the latter gets slated against the Planck scale giving rise to the well known discrepancy of proportion, $10^{120}$ orders of magnitude between the two \cite{weinberg}. This is what is caled the cosmological constant problem.

Leaving aside the question of vacuum energy, could we envisage some alternative way of generating  constant energy density, 
corresponding to the cosmological constant, $\Lambda$ ? One possible suggestion that has attracted attention is that the framework of General Relativity (GR) can be enlarged by having gravity propagate in extra compact space dimensions. In other words, the space has a microstructure. It turns out that by extending the framework of GR in this way we can generate the observed cosmological constant by tuning the size of the extra dimension to be of the order of $10^{-2}$ cm.

In what follows we envision that there is an extra compact dimension of radius $R$ sitting at each point in regular $3$-space which becomes visible only when we probe below this radius. It is generally believed in the spirit of string theory paradigm that all matter fields remain confined to the usual $4$-dimensional spacetime while gravity can propagate in  extra  dimensions. In this work, the extra  dimension is assumed to be compact and is limited  to a circle of radius $R$. Only gravity propagates in it. The fifth component of momentum is then quantized in units of $\frac{1}{R}$. Such a boundary condition generates a zero point energy (ZPE) which is proportional to $\frac{1}{2R}$ in the natural units for each mode via the quantum Uncertainity Principle. If it is only massless graviton, we know that it has two helicities (it is like a photon in a box which has two polarizations). Thus if $n$ is  number of extra dimensions with a degeneracy factor of 2, the quantum zero point energy (ZPE) is given by  
 
\begin{equation}
E_0  =   2n \frac{1}{2R}
\end{equation}

This ZPE  is a purely quantum effect that comes from compactifying the boundary of the extra dimensional space from infinity to a size/ radius $\sim R $. \\

\section{The energy density in the regular THREE dimensional space }

We use a simple argument to compute the energy density in the regular non-compact dimensions via  the ZPE due to compact dimensions. 
There are several works where extra dimensions have been invoked \cite{extra} to explain the cosmological constant. But in those works
it is the vacuum energy but not the zero point energy density that has been considered for such explanation. 
We consider here the topology of the space with one extra dimension as $R^3 \times S^1$. For simplicity we have assumed that each of our extra dimensions has the topology of $S^1$. As we zoom down in spatial resolution, we see the extra dimensions when we get to the  resolution scale, $R$, in the regular non-compact dimensions. 
Independent of the regular spacetime point we probe, the energy , $ E_0  =    \frac{n}{R} $ is sitting in a spherically symmetric $3$-volume,  $ V =   \frac{4\pi {R^3}}{3}$ of 
radius $R$. We choose such a volume to maintain the rotational symmetry of regular space. Remember that energy density is a local quantity. We get the energy density simply by dividing the zero point energy in this volume by the volume. This generates an energy density in the regular dimensions, 

\bea
\epsilon_0 =  \frac{A }{ R^4}  
\eea
where $A  = \frac{3n} {4 \pi}$.

 Thus we have a constant energy density all over homogeneous and isotropic space, then its conformity with the Einstein equation leads to the cosmological constant, $\rho=-p=\Lambda$. The ZPE in the compact dimension emerges as the cosmological constant in the physical $4$-dimensional Universe. 

\section{The nature of $\epsilon_0$ and negative pressure}

 Now, let us look a little more physically at the nature of this zero point energy density. To understand its effect  we will now start with a constant ground state energy density in the regular $3$-dimensional volume which is occupied by relativistic free quark matter and see 
 what kind of pressure it generates. \\

We consider a single quark flavour and massless quarks for simplicity. For quark matter in a given volume, let the energy density $\epsilon$, given below, be the sum of constant energy density, $B$, and  a matter part, where we have expressed the matter part as a function of the quark density\cite{baym}. It is given by

\bea
\epsilon     =   B   +  C {n_q}^{4/3} 
\eea
where  $ n_q =  \frac{k_f^3}{3\pi^{2}}$   is the number density of quarks, $ k_f$ is the Fermi momentum of the single flavour quark matter and $C$ is a phase space constant that depends on the degeneracy factor for quarks.\\

For quark matter the pressure is defined  with respect to the quark density as follows

\begin{equation}
p    =   n_q ^2 \cdot    \frac{d}{dn_q}(\epsilon /   n_q )=  - B  + (C/3) {n_q}^{4/3}
\end{equation}
Now we can immediately see the contribution of the  energy density, $B$,  to the pressure is exactly, $- B$.  This is what is required for the cosmological constant.  Actually, the quark matter was introduced as a device to define the pressure. Post calculation we can set the quark matter density to zero.

A good analogy is the MIT bag model \cite{mit} where the constant energy density, B,is used to mimic the `confinement mechanism' in QCD. In the case of the MIT bag an inward pressure, -B, is provided by the constant energy density, B. 

\section{ The usual Vacuum energy conundrum}

Though, we are not concerned with the usual quantum field theoretic vacuum energy density in this work, it may be appropriate to remark on this dilemma. The energy density corresponding to the cosmological constant is approximately

\begin{equation}
\epsilon_{\Lambda}  \sim  5 \times {10}^{-9}  \mathtt{erg/( cm^3 )} \sim 3 \times 10^3  \mathtt{eV/( cm^3 )}
\end{equation}

The usual quantum field theory vacuum energy  is calculated using closed vacuum energy Feynman diagrams which are bubble diagrams with no external legs. For example, in the scalar theory we have  the $ \phi^4$ interaction term which can generate two closed loops from one vertex. This has a quartic divergence that comes from two $4$ dimensional integrations and two closed propagators. It is usually regularized by dimensional regularization to remove the divergences and that leaves behind a quartic scalar mass term. These bubble diagrams do not contribute to any physical process - scattering etc. as they are without any external legs. Whether this vacuum energy will be observable physically is an open question. 

If we count in all the theories as we go up in energy scales, the hierarchy of theories leaves terms proportional to the fourth powers of all the masses in the theories. If we include masses up to the Planck mass scale, this yields a huge vacuum energy, that is  $ 10^{120} $ times the  observed cosmological constant energy density. This is the cosmological constant problem. \\

 For the evolution of the universe in the big bang scenario, the only quantum input is the Fermi -Dirac and Bose -Einstein distributions for fermions and bosons respectively. There is no field theoretic vacuum energy contribution. Let us now examine if the above contribution was there, how  would it change  nucleosynthesis.
The energy density of the universe corresponding to the higher temperarture $(~ 10^9 \; \mbox{K or} (100 \; \mbox{KeV} - 1 \; \mbox{MeV}))$ scale of nucleosynthesis is approximately

\begin{equation}
\epsilon_{N}  =   {10}^{22}  \mathtt{erg/( cm^3 )}
\end{equation}
Clearly  a vacuum energy contribution, $ 10^{120} $ times the the observed cosmological constant energy density will completely overpower the big bang nucleosynthesis energy density, despoiling  the accepted nucleosynthesis estimate. 
Such a large energy density would cause untenable expansion of the universe and disturb nucleosynthesis. Furthermore, it would lead to 
much lower CMB temperature. Thus the above vacuum energy contributions are physically inadmissible  because they contradict observed fact.

\section{ Simple estimate of the size of the extra dimension}

Several parameters of the universe has been determined by WMAP project \cite{wmap}. The Hubble constant $H_0$ is $0.7675 \times 10^{-28}$ cm$^{-1}$. The critical density of the universe is $\rho_c = \frac{3 H_0^2}{8 \pi G}\approx 5.313 \times 10^3$ eV cm$^{-3}$. The energy density
corresponding to cosmological constant $\Omega_\Lambda \approx 0.73 \pm 0.04$. This gives
$\Lambda = \rho_c \Omega_\Lambda \approx 3.96 \times 10^3 $ eV cm$^{-3}$.
The zero point  energy density in units of $\hbar c \approx 1.97 \times 10^{-5}$ eV cm in our case is  $ \epsilon_0 =  \frac{ 1.97 \times 10^{-5} A }{ R^4}   $. Comparing this with $\Lambda $ we get

\be     
\frac{ 1.97 \times 10^{-5} A }{ R^4}  \approx 3.96 \times 10^3 {\mbox{ eV cm}}^{-3}
\ee
which implies
\be
 R \approx 0.0084 \times  A^{1/4} \mbox{cm} 
\ee
$A$ as mentioned in Eq.(2) depends on $n$ the number of extra dimensions and for a few number of extra dimensions the required size of the extra dimension $R \lesssim 10^{-2}$ cm.

\section{ Discussion}

Let us first comment on the zero point energy density. It emanates from  compact dimension as the quantum mechanical energy difference between the unbounded  and the compact dimensions.  That is, it arises when the infinite extent of a dimension is confined to a finite size resulting in zero point energy. Its inclusion in homogeneous and isotropic FRW space-time defines the cosmological constant with $\rho=-p=\Lambda$. Further in contrast to the usual quantum field theory vacuum energy, it depends only on the size of the compact dimension and not on all the theories and masses thereof. Nor does it refer to any renormalization conditions which are ambiguous.



We shall end by remarking on the recent experiments  \cite{gravityexpt} that test Newton's law at small length scales of 100 $\mu$m. If we look at Newton's law at a scale less than R, then the extra dimension comes at par with regular dimensions and the gravitational force will go as  $ 1/(r^{2+n} )$ 
where $r<R$ and $n$ is number of extra dimensions. On the other hand if we look at length scales larger than R, we have the usual Newton's inverse square law operating. Therefore,  a direct probing of extra dimension would require measurement at sub millimetre scale. This is at the threshold of the present day experiments. What we have envisioned as a source for the cosmological constant could very well be tested observationally in the not too distant  future.\\ 

\begin{acknowledgments}
We like to thank M. Sami for discussion. 
\end{acknowledgments}

\end{document}